\documentclass{aa}
\usepackage{graphicx}
\usepackage{txfonts}
\begin{document}
\title{A method to search for topological signatures in the 
angular distribution of cosmic objects}

   \author{
Armando Bernui\inst{}\thanks{
on leave from Universidad Nacional de Ingenier\'{\i}a, 
Facultad de Ciencias, Apartado 31 - 139,\, Lima 31 -- Peru}
		  \and 
Thyrso Villela\inst{}
          }

   \offprints{A. Bernui}

   \institute{Instituto Nacional de Pesquisas Espaciais, 
   Divis\~{a}o de Astrof\'{\i}sica, Av. dos Astronautas 1758, 
   12227-010, S\~ao Jos\'e dos Campos, SP -- Brazil \\
\email{bernui@das.inpe.br} \\
\email{thyrso@das.inpe.br}             
             }

   \date{Received / Accepted}

\abstract{
We present a method to search for large angular-scale 
correlations, termed topological signatures, in the angular 
distribution of cosmic objects, which does not depend on 
cosmological models or parameters and is based only on the 
angular coordinates of the objects. 
In order to explore Cosmic Microwave Background temperature 
fluctuations data, we applied this method to simulated 
distributions of objects in thin spherical shells located 
in three dif\/ferent multiply-connected Euclidean 3-spaces 
($T^3$, $T_{\pi}$, and $G_6$), and found that the topological 
signatures due to these topologies can be revealed even if
their intensities are small.
We show how to detect such signatures for the cases of 
full-sky and partial-sky distributions of objects.
This method can also be applied to other ensembles of 
cosmic objects, like galaxies or quasars, in order to 
reveal possible angular-scale correlations in their 
distributions.
}

\titlerunning{Topological signatures in the angular 
distribution of cosmic objects
}
\authorrunning{Bernui \& Villela}

\maketitle

\section{Introduction} 

Homogeneity property refers to the spatial distribution of
cosmic objects in the Universe, and to test it one needs to
know the angular positions of these objects in the celestial 
sphere and their distances to us as well.
Unfortunately, cosmological distances are not directly 
measured, as they are calculated through the redshift-distance 
relationship, which involves cosmological parameters still 
not known with enough accuracy. 
So, precise distances to extragalactic objects are not 
currently available.
Isotropy, instead, deals with the angular distribution
of objects in the sky. Then, only two angular coordinates
for each object are needed to determine whether or not
they are isotropically distributed around us. 
Nowadays such coordinates are accurately measured and there 
are available catalogs for several types of objects, 
like quasars, gamma-ray bursts, and the (sky pixels  
corresponding to the) Cosmic Microwave Background Radiation 
(CMBR) temperature fluctuations. 
In analyzing such catalogs, some questions arise: 
How can we recognize whether a given class of cosmic objects 
follows an exact or a quasi-exact isotropic distribution? 
How can we quantify small deviations from a purely isotropic 
distribution? 
How can we distinguish a possible cosmological pattern or
topological signature imprinted in these catalogs from pure 
statistical fluctuations?

The study of the large-scale (i.e. global) homogeneity and 
isotropy of the Universe encompasses its topology as well, 
since topological properties are global properties of 
3-spaces. 
Multiply-connectedness is a topological property that 
tesselates a simply-connected space (like the Euclidean space 
${\cal R}^3$) producing multiple images of a given object. 
This property generates distance and angular correlations 
(i.e. anisotropies) in the distribution of objects.
The Small Universe hypothesis (Ellis \& Schreiber \cite{ES}),
which assumes that 
the Universe is a compact (i.e. multiply-connected with finite 
volume) 3-space that lies inside --at least partially-- the 
Last Scattering Surface (LSS), has not been discarded by 
recent CMBR data.
In view of this situation, we are motivated to study the 
anisotropic signatures of topological origin in the angular 
distribution of cosmic objects, and for this we consider 
simulated ensembles of objects in dif\/ferent Euclidean 
multiply-connected 3-spaces.

In order to search for these signatures in astronomical data 
sets, we developed a method based on histograms of angular 
separations between pairs of cosmic objects. 
We call it the Pair Angular Separation Histogram (PASH) 
method. 
We are particularly interested in studying angular 
correlations in simulated catalogs of cosmic objects 
that share the same attributes as the photons of the 
CMBR temperature fluctuations, i.e., point-like objects 
located in the (comoving 3-dimensional) thin shell 
representing the decoupling era, but viewed from 
Earth as located on the celestial sphere.
As we shall show, the correlations originated by the 
isometry properties of the compact 3-spaces can be 
revealed by the method we propose here. 
This method resembles the Cosmic Crystallography (CC)
method (Lachi\`eze-Rey \& Luminet \cite{LL}; 
Lehoucq et al. \cite{LLL}; 
Fagundes \& Gausmann \cite{FGa}, \cite{FGb}; 
Gomero et al. \cite{GRT1}, \cite{GRT2}, \cite{GRT3}, 
\cite{GTRB}).
However, there is a basic dif\/ference between them in 
that here we just look for angular correlations instead 
of searching for distance correlations as the CC method 
does.
Our method is therefore independent of any cosmological 
model or parameters, while the CC method depends on this 
information to calculate the radial distance of cosmic 
objects from their measured redshifts. 

In Sect. 2 we present the basic properties of the
3-spaces representing our Universe in the
Friedmann-Lema\^{\i}tre cosmological models
(Friedmann \cite{Friedmann}; Lema\^{i}tre \cite{Lemaitre}).
In Sect. 3, we present the geometrical and statistical 
method we developed to reveal correlations between pairs 
of objects distributed in a 2-dimensional spherical surface. 
The treatment of this problem initiates with the theoretical 
description of a perfectly isotropic distribution of objects 
in a spherical surface ${\cal S}^2$. 
For this type of data we construct the normalized expected
probability density for two objects be separated 
by a given angle, separation that of course ranges in the 
interval $[0,\pi]$.
This probability gives rise to the Expected Pair Angular
Separation Histogram (EPASH)\footnote{This result is 
independent of the three-dimensional geometry since all 
three isotropic geometries (i.e. spherical ${\cal S}^3$,
Euclidean ${\cal R}^3$, and hyperbolic ${\cal H}^3$) possess 
spherical surfaces ${\cal S}^2$ as hypersurfaces.}.
We also present the EPASH for partial-sky catalogs containing
objects distributed only in polar-caps.
In Sect. 4, we present the theoretical deduction of the
topological signatures due to pure translational isometries,
also known as Clifford translations, appearing in the PASHs.
We illustrate these results, in Sect. 5, with numerical 
simulations using catalogs that contain multiple images of 
cosmic objects generated by the isometries of the following 
Euclidean compact (i.e. multiply-connected with finite volume) 
orientable 3-spaces: $T^3$, $T_{\pi}$, and $G_6$
(denoted by ${\cal G}_1$, ${\cal G}_2$, and ${\cal G}_6$, 
respectively, in Wolf's classification Wolf~\cite{Wolf}). 
In this section we also show that performing the mean of 
several PASHs (or Mean Pair Angular Separation Histogram -- 
MPASH) reduces significantly the statistical noise, which 
allows us to reveal the anisotropies due to a non-translational 
isometry whose tiny signature is present in the angular 
distribution of cosmic objects.
We analized this problem considering MPASHs for full-sky
catalogs as well as for polar-cap catalogs.
Finally, in Sect. 6, we discuss our results and also
the problem of finding suitable catalogs of cosmic objects 
from the available astronomical data.

\section{The geometry of the 3-spaces}

The manifold $T^3$, best known as the three-torus, has its 
Fundamental Polyhedron (FP) usually --but not necesarily-- 
represented by an equal-sided cube,  
where its three pairs of faces are identified by pure 
(or Clifford) translation isometries.
The manifold $T_{\pi}$ has two pairs of faces identified by 
pure translation isometries and one pair of faces identified 
after a rotation of $180^{\circ}$, the so-called screw-motion 
isometry. 
The manifold $G_6$ has no pair of faces of its FP identified 
by a pure translation~\footnote{A picture of this FP can be 
seen, for instance, in Bernui et al. (\cite{BGRT}), where it 
is denoted by ${\cal T}_4$.}.
In analyzing the distributions of objects in these 3-spaces we 
shall be considering the effect of not only pure translational
isometries but also translation-with-rotation isometry, or 
screw-motion. 
For completeness, we find interesting to briefly review here 
the basic properties of compact 3-spaces of constant curvature.

The large-scale structure of the Universe is described by 
the Friedmann-Lema\^{\i}tre cosmological models.
In these models, the local geometry satisfies the local
homogeneity and isotropy properties and is described by the 
Robertson-Walker (RW) line element 
$ds^2 = dt^2 \,-\, a^2(t) d\sigma^2$, where $a(t)$ is the
scale factor, and 
$d\sigma^2 = d\chi^2 + {F_{k}}^2(\chi) \,
[ d\theta^2 + \sin^2 \!\theta d\varphi^2 ]$
is the line element of the $t=constant\,$ spatial sections.
The functions
$F_{k}(\chi) \equiv \sin\chi\/,\, \chi$\/, $\sinh\chi\,$, 
where $k$ is the normalized curvature parameter, which can 
take the values $+1, 0, -1$, corresponding to the spherical, 
Euclidean, and hyperbolic geometry, respectively.
It is known that the three-dimensional spatial sections
(hereafter denoted by ${\cal M}_k$) are manifolds of constant
curvature obtained from the quotient 
${\cal M}_k = \widetilde{{\cal M}}_{k} / \Gamma$
(see e.g. Wolf~\cite{Wolf}), where $\Gamma$ is a discrete group 
of isometries of the manifold $\widetilde{{\cal M}}_{k}$, 
without fixed points. 
The manifold $\widetilde{{\cal M}}_{k}$ is termed the 
universal covering of ${\cal M}_k$.
Therefore, while $\widetilde{{\cal M}}_{k}$ is a 
simply-connected manifold, ${\cal M}_k$ is a multiply-connected 
one.
We would like to stress that the values $k\,= \,+1, 0, -1$ in
the RW line element determines only the local geometry of
${\cal M}_k$ (equivalently that of $\widetilde{{\cal M}}_{k}$).

Since the simply-connected $\widetilde{{\cal M}}_{k}$ and the
multiply-connected ${\cal M}_k$ manifolds share the assumed
local homogeneity and isotropy properties, we conclude that
both manifolds are good solutions of the RW local geometry, 
i.e., the Einstein-Hilbert equations. 
Consequently, General Relativity (or any other metrical theory)
can not distinguish between $\widetilde{{\cal M}}_{k}$ and
${\cal M}_{k}$ because they look like each other locally.

In a Universe with a multiply-connected 3-space ${\cal M}_k$, 
and whenever the horizon scale is greater than half of the 
smallest closed geodesic of ${\cal M}_k$, multiple images 
of a given cosmic object might exist due to the isometries 
of the group $\Gamma$.\footnote{Except when $\Gamma=\mbox{Identity}$, 
in which case ${\cal M}_k = \widetilde{\cal M}_k$ is a 
simply-connected manifold.} 
These images originate distance correlations between the 
mapped cosmic objects and also produce angular 
correlations between those distance-correlated objects. 
As we shall see, these angular correlations can be revealed 
by the PASH method described in the next section.

\section{PASHs}
\subsection{EPASHs for full-sky catalogs}

In essence, a PASH is a normalized plot of the number of pairs
of objects versus the angular distance between them.
A formal construction of a PASH is as follows:
let ${\cal B}_R \subset {\cal M}_k$ be a ball of radius $R$
centered at the origin of coordinates ${\cal O}$, and 
involving spherical surface ${\cal S}_{R}$, and containing an
ensemble of (cosmic) objects.
We call ${\cal B}_R$ the observable Universe, ${\cal O}$ 
represents the Earth, $R_{-}$ is the radius of 
the LSS, ${\cal S}_R$ is the celestial sphere with radius 
$R \equiv R_{+} = R_{-} + \Delta R$, where $\Delta R$ is 
the thickness of the decoupling era. 
Since we are interested in analyzing CMB data, our 
simulations consider ensembles of objects located in a thin 
shell of thickness $\Delta R$, which are after projected 
onto the celestial sphere ${\cal S}_R$ ignoring their radial 
coordinate. 
However, we point out that this method can be applied to 
any spatial distribution of cosmic objects.
Given a set of well-defined selection rules\footnote{i.e. 
physical properties that are common to the cosmic objects in 
a given catalog, such as luminosity threshold, redshift, etc.}, 
a catalog ${\cal C}$ is a list of all objects in ${\cal B}_R$ 
which satisfies a sub-set of the whole set of rules.

We start the description of the method by firstly showing
how to obtain the PASH.
Let's assume that we have an ensemble of objects that are, 
by assumption, isotropically distributed in ${\cal S}_R$
(see Eq.~\ref{iso_dis}).
A full-sky map catalog ${\cal C}$ is a list of $N$ objects 
spreaded out in ${\cal S}_R$ with their corresponding angular 
coordinates.
We divide the interval $(0,\pi]$ in $m$ bins of equal length
$\delta \gamma \,=\, \pi / m$, where each sub-interval has the
form $J_i = (\gamma_{i}^{} - \frac{\delta \gamma}{2} \, , \,$
$\gamma_{i}^{} +\frac{\delta \gamma}{2}] \,,\, 
i= 1,2, \dots ,m \, ,$ with center in
$\gamma_{i}^{} = \,(i - \frac{1}{2}) \,\, \delta \gamma \,$.
Now, we denote by $\phi(\gamma)$ the number of pairs of objects
in ${\cal C}$ separated by a distance $\gamma \in (0,\pi]$.
Thus,
\begin{equation} \label{PASH0}
\Phi(\gamma_{i}^{}) = \frac{2}{N(N-1)\,\delta \gamma} \,\,
\sum_{\gamma \in J_i} \phi(\gamma)
\end{equation}
is the normalized counting of the number of pairs of objects,
separated by an angle $\gamma_{i}^{}$, that lies in the
sub-interval $J_i$. Actually,
$\sum_{i=1}^m \Phi(\gamma_{i}^{})\, \delta \gamma = 1 \,$, 
so we can average several functions $\Phi(\gamma_{i}^{})$,
each one built from catalogs having a comparable number 
of objects.

It may occur that some angular correlations of small intensity
are present in a given PASH, but are not clearly seen.
To reveal them, and since PASH and EPASH are normalized, 
we subtract the PASH from the EPASH in order to enhance 
possible angular correlations.

Let us now obtain the EPASH for the simply-connected manifold 
$\widetilde{{\cal M}}_k$.
Let $p, q \in {\cal B}_R \subset \widetilde{{\cal M}}_k$ be 
an arbitrary pair of objects listed in ${\cal C}$ with 
coordinates $(r_p,\theta_p,\varphi_p)$, $(r_q,\theta_q,\varphi_q)$,
respectively.
Then, the probability density that two completely 
uncorrelated objects$\:$\footnote{this situation corresponds 
to the monopole term ($l = 0$) in a multipole decomposition.} 
$p$ and $q$ be separated by an angle $\gamma \in (0,\pi]$, is
\begin{equation} \label{PASH}
{\cal P}(\gamma)
\!=\!\! \int_{{\cal B}_R} \!\! \int_{{\cal B}_R} 
\!\!\! d^3r_{p} d^3r_{q} \rho({\,\bar{\!r}}_p) 
\rho({\,\bar{\!r}}_q)\,\delta(\, \Lambda(p,q) - \gamma \,) \,,
\end{equation}
where ${\,\bar{\!r}}_p, {\,\bar{\!r}}_q$ are the vectors 
from ${\cal O}$ to the position of the objects $p$ and $q$,
respectively.
In $\widetilde{{\cal M}}_0 = {\cal R}^3$, the angular
separation between $p$ and $q$ is given by
\begin{equation}
\Lambda(p,q) \equiv \arccos[\cos \theta_p \cos \theta_q + 
\sin \theta_p \sin \theta_q \cos(\varphi_p - \varphi_q)],
\end{equation}
where $\theta_p,\, \theta_q \in [0,\pi]$ and 
$\varphi_p,\, \varphi_q \in [0,2\pi]$.
Note that $\rho({\,\bar{\!r}}_p) d^3r_{p}$ is the 
probability of the object $p$ to be in the volume $d^3r_p$.
The probability of finding $p$ in the whole ${\cal B}_R$
should of course be equal to 1, i.e.,
\begin{equation}
\int_{{\cal B}_R}\,\, \rho({\,\bar{\!r}}_p) \, d^3r_p = 1\, .
\end{equation}
Thus, a purely isotropic and normalized density distribution
of objects observed in ${\cal S}_R$ reads
\begin{equation} \label{iso_dis}
\rho({\,\bar{\!r}}_p) = \frac{\delta(r_p - R)}{4 \pi R^2} \, .
\end{equation}
Finally, integrating~(\ref{PASH}) we obtain
\begin{equation} \label{EPASH}
{\cal P}(\gamma) = \, \frac{1}{2} \, \sin \gamma \,\, .
\end{equation}
Hence, the EPASH is simply defined by
\begin{equation} \label{def_expected}
\Phi_{expected}(\gamma_{i}^{}) \equiv 
\frac{1}{\delta \gamma} \,
\int_{J_i} {\cal P}(\gamma) \, d\gamma \, .
\end{equation}
However, if the interval $J_i$ is small enough (as shall be
considered here), a suitable approximation for the EPASH is
\begin{equation} \label{EPASHFinal}
\Phi_{expected}(\gamma_{i}^{}) \simeq 
{\cal P}(\gamma_{i}^{})\, .
\end{equation}
\noindent
We also note that ${\cal P}(\gamma)$ defined in 
Eq.~(\ref{PASH}) satisfies the normalization property,
$\int_{0}^{\pi} {\cal P}(\gamma) \, d\gamma \,=\, 1$,
which is a useful condition since it let us to perform the 
mean of an arbitrary number of histograms obtained with 
catalogs containing a comparable number of objects.

\begin{figure}
\includegraphics[width=9cm]{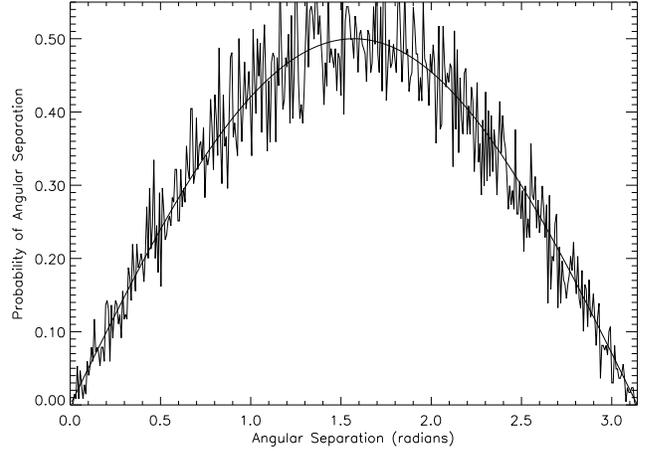} 
\caption{EPASH (smooth curve) and PASH (noisy curve)
for a full-sky simulated catalog with $N\,=\,200$ objects 
in ${\cal S}_R;\, m=400$ bins.}
\label{figure1}
\end{figure}

\begin{figure}
\includegraphics[width=9cm]{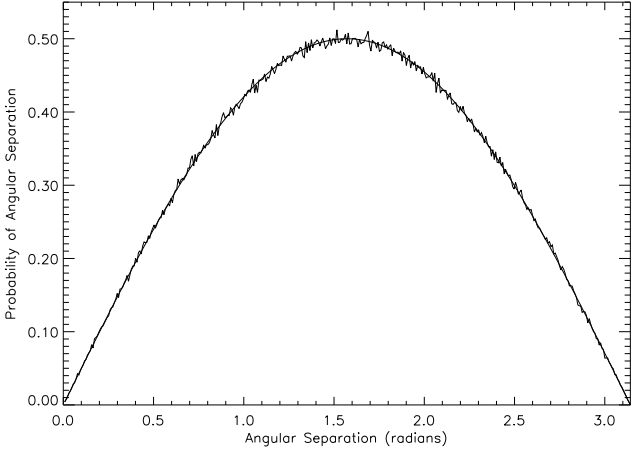}
\caption{EPASH (smooth curve) and PASH (noisy curve) for a 
full-sky simulated catalog with $N\, =\,2 \times 10^3$ 
objects in ${\cal S}_R;\, m=400$ bins.} 
\label{figure2}
\end{figure}

In Figs.~\ref{figure1} and~\ref{figure2} we show the 
EPASH together with the corresponding PASHs obtained 
from numerical simulations for full-sky catalogs with 
$N\, =\, 200$ and $N\, =\, 2 \times 10^3$ objects, 
respectively. 

\subsection{EPASHs for polar-cap catalogs}

Our galaxy poses a serious problem to the use of full-sky 
CMBR maps and for other sky catalogs as well, because it 
contaminates part of the celestial sphere around the 
Galactic plane.
This fact makes somewhat useless the approach we developed 
studying the angular distribution of cosmic objects listed 
in full-sky catalogs. For this reason, we studied also the 
PASHs for catalogs containing objects located in polar-caps 
(i.e. above or below the equator of a spherical distribution).

Consider a catalog with objects located in a spherical cap
surface $S_{PC} \subset {\cal S}^2$, centered, for instance,
around the $z$-axis, which is assumed to be aligned with the
North Pole. We call this ensemble a polar-cap catalog 
${\cal C}_{PC}$.
In this case, any object with coordinates
\begin{eqnarray}
x\,&=&\, R\,\cos\varphi\,\sin\theta \, , \nonumber \\
y\,&=&\, R\,\sin\varphi\,\sin\theta \, , \\
z\,&=&\, R\,\cos\,\theta \, , \nonumber
\end{eqnarray}
where $\varphi \in [0,2\pi]$ and the azimuthal angle
$\theta \in [0,\theta_0]$, with $\theta_0 \leq \pi/2$,
is in the surface $S_{PC}$.
Thus, the isotropic and normalized density distribution of
objects in $S_{PC}$ is
\begin{eqnarray}
\rho({\,\bar{\!r}}_p)
= \frac{\Theta(\theta_0 - \theta_p) \, \delta(r_p - R)}
{A_{PC}} \, ,
\end{eqnarray}
where $A_{PC} \equiv 2\,\pi\,R^2\,(1 - \cos\theta_0)$ 
is the area of the polar-cap.
Using this information in the definition of an EPASH
(Eq.~\ref{PASH}) and after the integration of the 
$\delta$ Dirac-function, we obtain
\begin{eqnarray} \label{EPASHpc}
\!\!{\cal P}_{PC}(\gamma;\theta_0)\!= \!\! \frac{R^4}{A_{PC}^2}
\!\! \int_0^{\theta_0} \!\!\!\!\!\!\sin\theta_p d\theta_p
\!\! \int_0^{\theta_0} \!\!\!\!\!\!\sin\theta_q d\theta_q
I(\theta_p,\theta_q;\gamma) , 
\end{eqnarray}
where
\begin{eqnarray}
I(\theta_p,\theta_q;\gamma) \equiv
\int_0^{2\pi} \! \int_0^{2\pi} \!\! d\varphi_p d\varphi_q
\delta(\Lambda(p,q) - \gamma) . 
\end{eqnarray}
It can be shown that this integral is equal to
\begin{eqnarray}
I(\theta_p,\theta_q;\gamma) =
\frac{4 \pi \sin\gamma}{\sin\theta_p\,\sin\theta_q\,
\sin\alpha(\theta_p,\theta_q;\gamma)} \, ,
\end{eqnarray}
where the function $\alpha$ is defined by
\begin{eqnarray}
\alpha(\theta_p,\theta_q;\gamma) \equiv
\arccos\left[\frac{\cos\gamma - \cos\theta_p \cos\theta_q}
{\sin\theta_p\sin\theta_q}\right] \, 
\end{eqnarray}
and lies in the interval $[0,\pi]$.
Due to this condition, $\theta_p$ and $\theta_q$ should 
satisfy the relation 
\begin{eqnarray} 
\frac{\cos\gamma - \cos\theta_p \cos\theta_q}
{\sin\theta_p\sin\theta_q}
\, \in \, [-1,1] \, .
\end{eqnarray}
Equivalently, this constraint for the values of $\theta_p$ 
and $\theta_q$ can be written as the following two 
inequalities:
\begin{eqnarray} \label{conditions} 
\theta_p + \theta_q \, &\geq& \gamma \, , \\
|\theta_p - \theta_q| &\leq&
\mbox{\rm min}\{\theta_0,\gamma\}\, . \nonumber
\end{eqnarray}
Thus, the probability density to observe two 
objects in a polar cap (defined by the angle $\theta_0$) 
with an angular separation $\gamma$ is
\begin{eqnarray} \label{EPASHpc}
\!{\cal P}_{PC}(\gamma;\theta_0)
\!=\! \frac{4\pi R^4 \sin\gamma}{A_{PC}^2} \!\!
\int_0^{\theta_0} \!\!\! d\theta_p \!\!
\int_0^{\theta_0} \!\!\!\!\! 
\frac{d\theta_q}{\sin\alpha(\theta_p,\theta_q;\gamma)} \, ,
\end{eqnarray}
with the conditions given in Eq.~(\ref{conditions}). 
The double integral in Eq.~(\ref{EPASHpc}) can be 
numerically solved, but it is convenient to do it by 
parts considering carefully the conditions imposed by the 
inequalities in Eq.~(\ref{conditions}).
Thus, given $\theta_0$, we can obtain the probability 
density ${\cal P}_{PC}$ as a function of the 
angular separation $\gamma$.

\begin{figure}
\includegraphics[width=9cm]{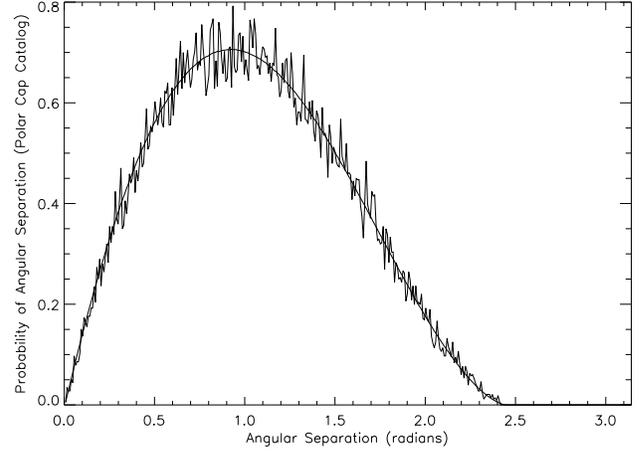}
\caption{EPASHpc (smooth curve) and PASH (noisy curve) for a 
polar-cap simulated catalog with $\theta_0 = 70^{\circ}$, 
$N_{PC}\, =\, 332$ objects ($10^3$ objects in the full-sky 
simulated catalog, $m = 400$ bins); $\sigma = 0.02799$.} 
\label{figure3}
\end{figure}

\begin{figure} 
\includegraphics[width=9cm]{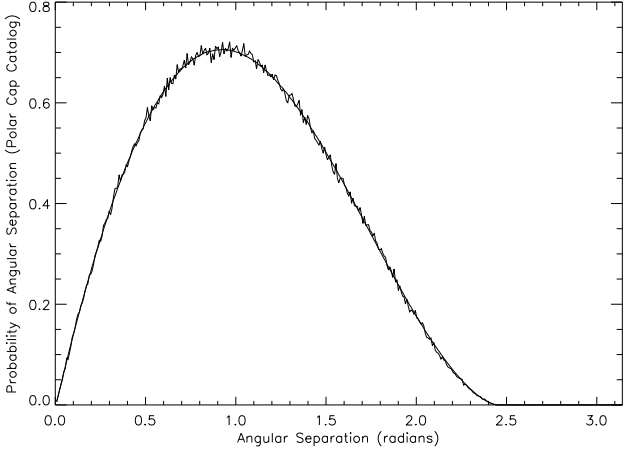}
\caption{EPASHpc (smooth curve) and PASH (noisy curve)
for a polar-cap simulated catalog with $\theta_0 = 70^{\circ}$, 
$N_{PC}\, =\, 1337$ objects (with $4 \times 10^3$ objects in 
the full-sky simulated catalog, $m = 400$ bins); 
$\sigma = 0.00661$.}
\label{figure4}
\end{figure}

Recently, the analytical expression for the expected 
probability density that an arbritary pair 
of objects in $S_{PC}$ be separated by $\gamma$ degrees, 
i.e. EPASH for polar-caps (EPASHpc), has been found by 
Teixeira (\cite{AT}), who obtained 
\begin{eqnarray} \label{EPASHpcAT}
{\cal P}_{PC}(\gamma;\theta_0) \, = \, 
\frac{4\pi R^4 \sin\gamma}{A_{PC}^2} 
\left( \, \arcsin[\,\{(\cos \theta_0 + \cos \gamma) \right.  
&& \nonumber \\
\sqrt{\cos \gamma - \cos 2\theta_0}\, \} \,
/ \, \{ \sin \theta_0\, (1 + \cos \theta_0)\,\sqrt{1+\cos \gamma} 
\, \} \,] && \nonumber  \\
\left. + \,\,(1 - 2\,\cos \theta_0)\,
\arccos[\, \cot \theta_0\, \tan (\gamma/2) \,]\, 
\right) . &&  
\end{eqnarray}
To illustrate this result, we simulate two full-sky 
catalogs with $N\, =\, 10^3$ and $N\, =\, 4 \times 10^3$ 
objects, respectively.
We show in Figs.~\ref{figure3} and~\ref{figure4} the 
EPASHpc together with the corresponding PASHs obtained 
from these simulations, where the number of objects in the 
polar-caps were $N_{PC}\, =\, 332$ and $N_{PC}\, =\, 1337$, 
respectively.

\subsection{MPASHs for full-sky and polar-cap catalogs}

Suppose that $K$ comparable catalogs ${\cal C}_k, \,k=1,2,...,K$, 
of objects located in the celestial sphere are available for 
analysis.
For a given $m$ (common to all the PASHs), we calculate the
functions
\begin{equation}
\Phi_k(\gamma_{i}^{})= \frac{2}{N_k(N_k-1)\,\delta \gamma} \,\,
\sum_{\gamma \in J_i} \phi_k(\gamma) \, ,
\end{equation}
where $N_k$ is the number of objects in ${\cal C}_k$, and
$\phi_k(\gamma)$ is the number of pairs of objects in
${\cal C}_k$ with angular separation $\gamma$.
Then, the MPASH is defined by 
\begin{equation} \label{MPASH}
\left< \Phi(\gamma_{i}^{}) \right> \,\, \equiv \,\frac{1}{K} \,
\sum_{k=1}^{K} \Phi_k(\gamma_{i}^{}) \, .
\end{equation}
Indeed, we know that performing the average of Pair Separation 
Histograms in the CC method, the result is that the statistical 
fluctuations are reduced by a factor proportional to $1/ \sqrt{K}$ 
but the topological information is preserved 
(Gomero et al. \cite{GRT1}, \cite{GRT2}, \cite{GRT3}). 
In the next section we shall see that for the PASHs a similar 
situation occurs, that is, the greater the number of PASHs 
considered in the average, the better the approximation of the 
MPASH to the corresponding EPASH.

\subsection{Noise in PASHs}

\begin{figure}  
\includegraphics[width=9cm]{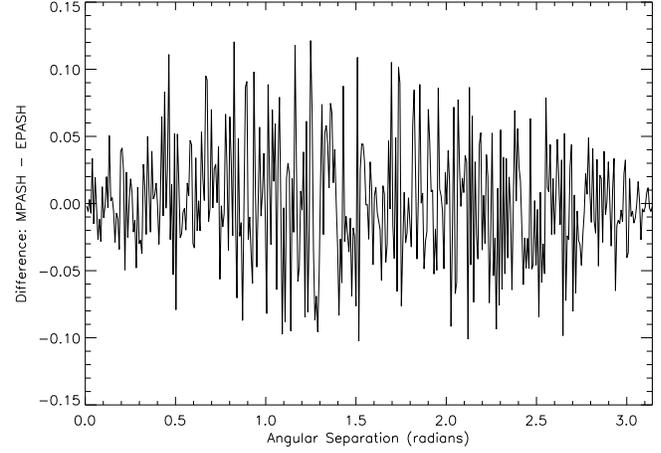}
\caption{Dif\/ference between PASH and EPASH plotted in 
Fig.~\ref{figure1} for a full-sky simulated catalog ($K=1$) 
with $N=200$ objects and $m = 400$ bins; $\sigma= 0.04418$.}
\label{figure5}
\end{figure}

\begin{figure}  
\includegraphics[width=9cm]{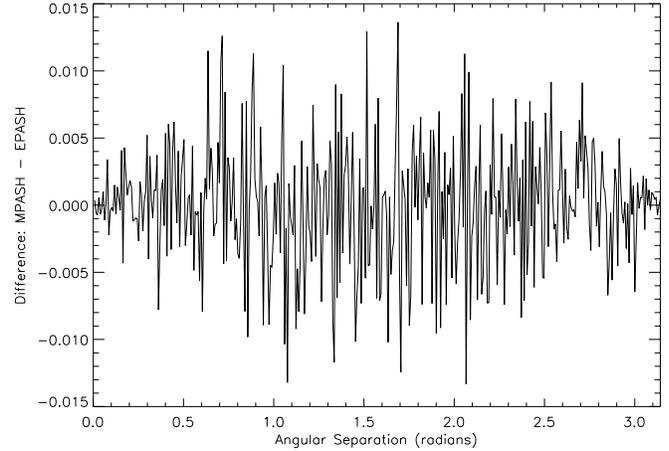}
\caption{Dif\/ference between PASH and EPASH plotted in 
Fig.~\ref{figure2} for a full-sky simulated catalog ($K=1$) 
with $N=2 \times 10^3$ objects and $m = 400$ bins; 
$\sigma= 0.00453$.
Note that the scale here is dif\/ferent from that of 
Fig.~\ref{figure5}.}
\label{figure6}
\end{figure}

In Figs.~\ref{figure5} and~\ref{figure6} we show the 
dif\/ference between the PASH and the EPASH, already plotted 
in Figs. 1 and 2, respectively.
The noise of a PASH, as observed in Figs.~\ref{figure5} 
and~\ref{figure6}, is simply given by 
\begin{equation}
\eta(\gamma_{i}^{}) = \Phi(\gamma_{i}^{}) 
- \Phi_{expected}(\gamma_{i}^{}) \, ,
\end{equation}
and is obtained from the very def\/inition of the EPASH.
In fact, given a catalog $\cal{C}$ with $N$ objects, we 
can quantify the statistical noise in a PASH by:
\begin{eqnarray} \label{sigma_noise}
\sigma \equiv \sqrt{\frac{1}{m} \, \sum_{i=1}^{m} \, 
\eta(\gamma_{i}^{})^2 \,} \,\, .
\end{eqnarray}
For the cases shown in Figs.~\ref{figure5} and~\ref{figure6}, 
we evaluate the statistical noise using Eq.~(\ref{sigma_noise}) 
obtaning $\sigma= 0.04418$ for $N=200$, and 
$\sigma = 0.00453$ for $N=2 \times 10^3$ respectively. 

\begin{figure} 
\includegraphics[width=9cm]{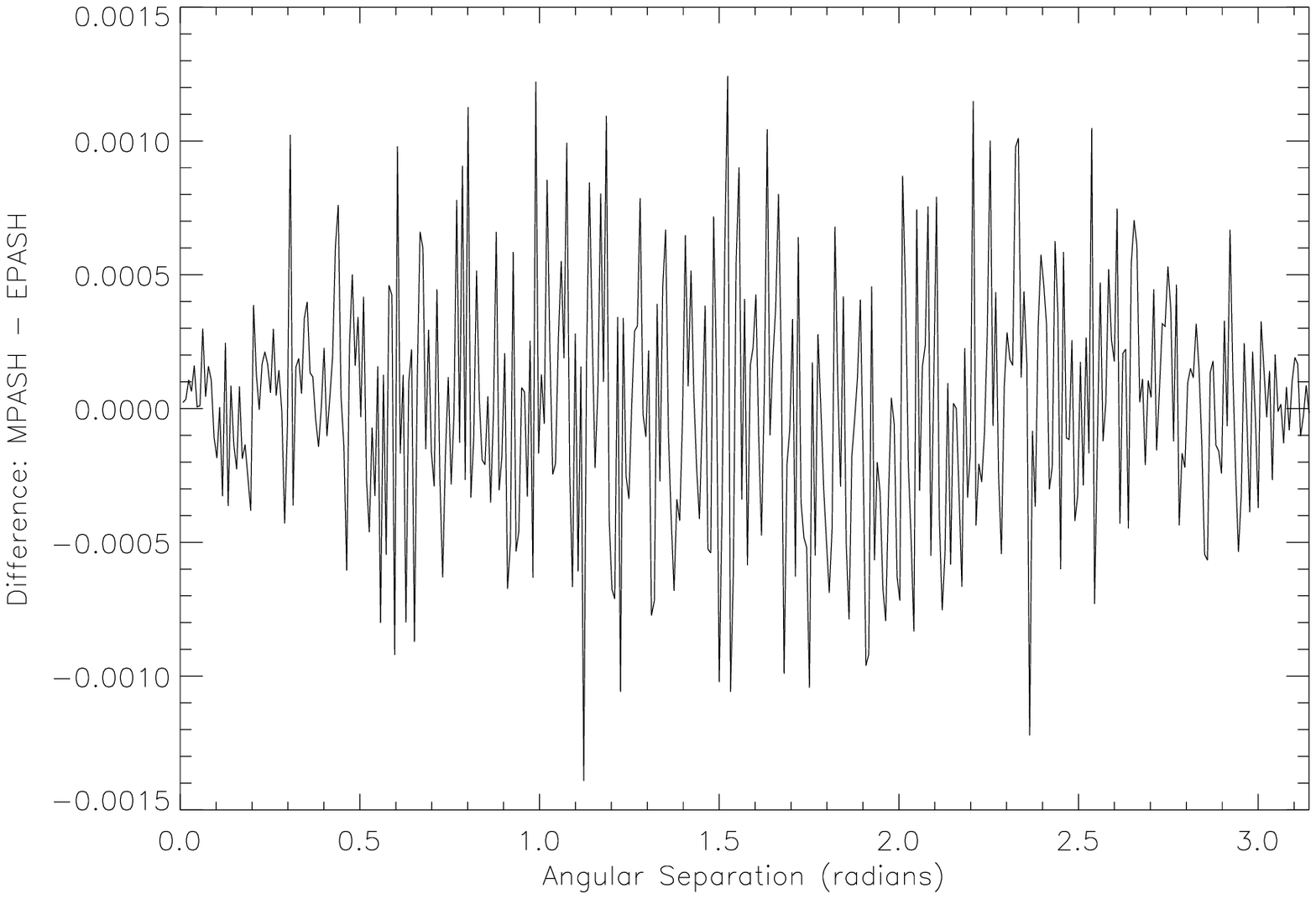}
\caption{Dif\/ference between MPASH (for a 
numerical simulation of $K= 10^4$ full-sky simulated 
catalogs in ${\cal R}^3$, with 
$\left< N \right> \,=\,200$ objects and $m = 400$ bins) 
and EPASH; $\sigma= 0.00045$.} 
\label{figure7}
\end{figure}

\begin{figure} 
\includegraphics[width=9cm]{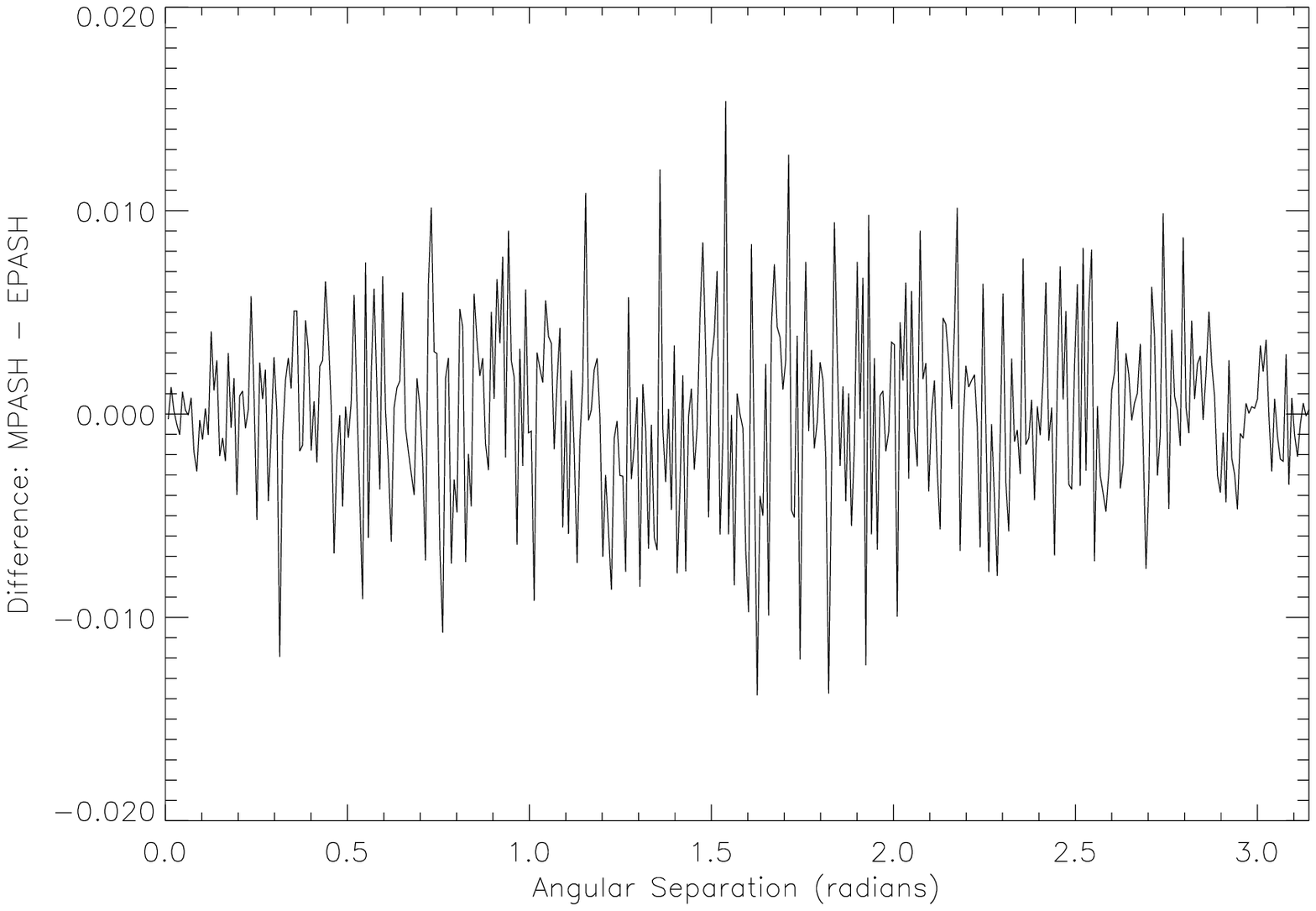}
\caption{Dif\/ference between MPASH (for a numerical 
simulation of $K= 10^4$ full-sky simulated catalogs in 
${\cal R}^3$, with $\left< N \right> \,=\,20$ objects and 
$m = 400$ bins) and EPASH;
$\sigma=0.00448$.}
\label{figure8}
\end{figure}

\begin{figure} 
\includegraphics[width=9cm]{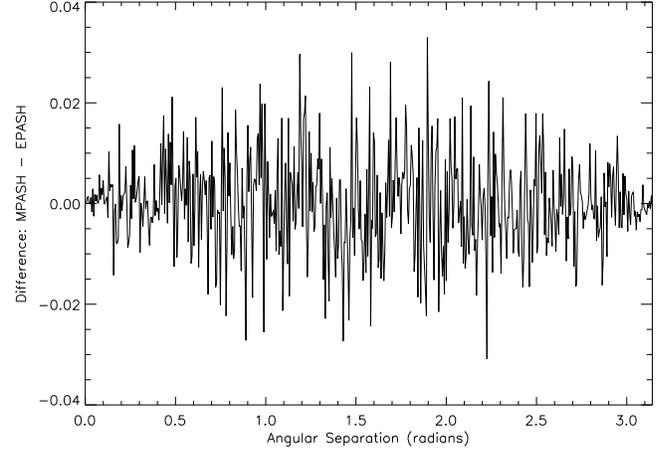}
\caption{Dif\/ference between MPASH (for a numerical 
simulation of $K=150$ full-sky simulated catalogs in 
${\cal R}^3$, with $\left< N \right> \,=\,100$ objects 
and $m = 600$ bins) and EPASH; $\sigma=0.00926$.}
\label{figure9}
\end{figure}

In Fig.~\ref{figure7}, we show the dif\/ference between the 
MPASH and the EPASH for a numerical simulation of $K= 10^4$ 
full-sky simulated catalogs in ${\cal R}^3$, with a mean 
number of 
$\left< N \right> \equiv 1/K \sum_{i=1}^{K} N_i \,=\,200$
objects per catalog. 
The statistical noise is $\sigma=0.00045$. 
Fig.~\ref{figure8} shows the dif\/ference between a MPASH, 
obtained from a numerical simulation of $K= 10^4$ full-sky 
catalogs with $\left< N \right> \,=\,20$ objects each, 
and the corresponding EPASH; in this case $\sigma=0.00448$.
In Fig.~\ref{figure9}, we show the dif\/ference between a 
MPASH and the EPASH, where the MPASH was obtained from a 
numerical simulation of $K=150$ full-sky simulated catalogs 
with $\left< N \right> \,=\,100$ objects each, but with 
$m=600$. In this case $\sigma=0.00926$. 

Thus, when we compare the cases shown in Figs.~\ref{figure6} 
and~\ref{figure8} (equal $m$ and $\sigma$, dif\/ferent values 
for $N$ and $K$) with the cases shown in Figs.~\ref{figure5} 
and~\ref{figure7} (equal $m$ and $N$, dif\/ferent values for 
$\sigma$ and $K$), we finally arrive at
\begin{equation} \label{sigma}
\sigma \simeq 
\frac{0.45\,\sqrt{m}}{\left< N \right> \,\sqrt{K} } \, .
\end{equation}
We can verify this relationship with the aid of the case 
shown in Fig.~\ref{figure9}, in which $m=600$. 
In fact, using Eq.~(\ref{sigma}), we obtain 
$\sigma=0.009$, while calculating the right hand side of 
Eq.~(\ref{sigma_noise}) with the values of 
$\Phi(\gamma_{i}), \, i=1,...,600$, obtained from 
Fig.~\ref{figure9}, results $\sigma=0.00926$. 

In Figs.~\ref{figure5} and~\ref{figure6}, as well as in 
Figs.~\ref{figure3} and~\ref{figure4}, we notice the 
absence of correlations between pairs of objects for the 
exactly isotropic distribution in ${\cal R}^3$ generated 
using random numbers for the coordinates.
Thus, it is clear from these simulations that as
$K \rightarrow \infty$ and/or $N \rightarrow \infty$ the
statistical noise measured by $\sigma \rightarrow 0$.
The MPASH is related to the statistical noise and this fact 
is quantified in Eq.~(\ref{sigma}).
Although the statistical noise is inherent to any numerical 
simulation, it can be controlled in three ways: i) by 
obtaining catalogs with a large number of elements, in order
to produce PASHs with a small noise, since the statistical 
noise depends on the number of objects;
ii) by computing the MPASH to significantly reduce the 
statistical noise;
iii) by choosing a suitable number of bins $m$.

\section{Topological signatures in PASHs}

In any multiply-connected 3-space ${\cal M}_0$ whose FP lies 
inside the ball ${\cal B}_R$, there might exist images of a 
given object in a shell of thickness 
$\Delta R \equiv R - R_{-}$, whenever the horizon scale 
$R$ is greater than half of the length of the smallest 
closed geodesic of ${\cal M}_0$ 
(Lachi\`eze-Rey \& Luminet \cite{LL}; 
Lehoucq et al. \cite{LLL}; 
Fagundes \& Gausmann \cite{FGa}, \cite{FGb}; 
Gomero et al. \cite{GRT1}, \cite{GRT2}, \cite{GRT3}, 
\cite{GTRB}).
These images give extra contributions to the PASH due to
the isometries of ${\cal M}_0$.

By definition, the probability density for two 
objects $p, \, q$ in ${\cal C}$ be correlated by a 
translational isometry $g_i$ (we represent this correlation 
by $q=g_{i}(p)$\,) is 
\begin{eqnarray} \label{Isometry_def}
{\cal P}^{g_{i}}(\ell) \equiv \frac{\nu_{g_{i}}}{n}
\int_{{\cal B}_R} \int_{{\cal B}_R} 
d^3r_{p} d^3r_{q} \rho({\,\bar{\!r}}_p) \rho({\,\bar{\!r}}_q) 
\times \nonumber \\  
\delta(d(p,q) - \ell)\,
\delta({\,\bar{\!r}}_q -g_{i}({\,\bar{\!r}}_p))\, ,
\end{eqnarray}
where $\ell$ is the distance variable for objects in 
${\cal B}_R$, $\ell \in (0,2R]$; 
$d(p,q) \equiv |{\,\bar{\!r}}_p - {\,\bar{\!r}}_q|$
is the 3-dimensional Euclidean distance between $p$ and $q$,
and the coef\/ficients $\nu_i \equiv \nu_{g_{i}}$ are the 
number of images generated by the isometry $g_i$.
Moreover, $n$ is the density number of objects inside the
volume under investigation (in the present case a thin shell), 
and $N$ is the number of objects in this volume.

Consider the pure translational isometries in ${\cal M}_0 = T^3$, 
with a FP represented by a cube of side $L$.
We know that the distances between pairs correlated by Clifford 
translations are independent of the location of the 
object (see Gomero et al. \cite{GTRB}), i.e., 
$\forall \, p \in {\cal B}_R$, the distance 
$|g_i(p)| = d(p,g_i(p)) = \lambda_i$ is independent of $p$. 
Thus, for this manifold, the translational isometries 
$g_i, \, i=1,2,...$ are such that for $\lambda_{\,1} = L$ 
we obtain $\nu_1 = 6$, for $\lambda_{\,2} = \sqrt{2}L$ we 
obtain $\nu_2 = 12$, for $\lambda_{\,3} = \sqrt{3}L$ we 
obtain $\nu_3 = 8 $, etc.

Let's now apply the above definition to the topological 
images correlated by the Clifford translation isometries 
$g_i$ and located in a thin shell of thickness $\Delta R$.
In this case, 
\begin{eqnarray}
\rho({\,\bar{\!r}}_p)= \frac{\Theta(R - r_p)\, 
\Theta(r_p - R_{-})}{\frac{4 \pi}{3} [R^3 - R_{-}^3]} \, . 
\end{eqnarray}
Using this expression in Eq.~(\ref{Isometry_def}), and 
after some calculations, we arrive at 
\begin{eqnarray} \label{TI}
{\cal P}^{g_i}(\ell) = P_i \, \Theta(R - \frac{\lambda_i}{2}) 
\, \delta(\ell - \lambda_i) \, , 
\end{eqnarray}
which, due to the $\delta$ Dirac-function, appears as an acute 
peak localized at $\ell = \lambda_i$ in any pair 
distance-separation histogram, and where 
\begin{eqnarray} \label{P_i}
P_i \equiv \frac{3\,\nu_{i} \, \Delta R 
\,(R + R_{-})^2}
{8\, \lambda_i \,N\, (R^2 + R R_{-} + R_{-}^2)} 
\end{eqnarray}
gives the height of the peak.
We notice that the probability for these isometries to show up 
in pair separation histograms is proportional to the thickness 
$\Delta R$ of the shell and inversely proportional to $N$, the 
number of objects in the shell.

In Euclidean multiply-connected 3-spaces with Clifford 
translations, the objects located in a thin shell and 
correlated in distance by the Clifford translational isometries 
$g_i$ (of the covering group $\Gamma$) give rise to a definite 
signature in the PASHs. 
In fact, for shells of small thickness (like that corresponding 
to the decoupling era where $\Delta R/R \simeq 0.003$), these 
angular correlations between pairs appear in the MPASHs as 
small peaks and are approximately located at the angular scales 
\begin{eqnarray}
\gamma_{g_i}^{} \, \simeq \, 2 \,
\arcsin \left( \frac{\lambda_{\,i}}{2\,R} \right) \, ,
\end{eqnarray}
whenever $R \ge \lambda_{min}/2$, which is the requirement
for the existence of topological images in the observable 
universe.
Thus, it is not dif\/ficult to verify that the minimum and the
maximum angle ($\gamma^{min}_{g_i}$ and $\gamma^{max}_{g_i}$, 
respectively) subtended by two objects located in a thin 
shell and separated by a isometry distance $\lambda_{\,i}$ 
are, respectively:
\begin{eqnarray}
\sin(\gamma^{min}_{g_i}/2)\!\!
& \simeq &\!\!\frac{\lambda_i}{2\,R}\, , \\
\sin(\gamma^{max}_{g_i}/2)\!\!
& \simeq &\!\!\frac{\lambda_{\,i}}{2\,R_{-}}\, .
\end{eqnarray}
We observe that the probability for angular correlation 
between objects is a direct consequence of the existence 
of the $N(N-1)/2$ pairs located in the thin shell and 
correlated by the Clifford translation $\lambda_i$ 
with the probability given in Eq.~(\ref{P_i}).
In fact, the probability density that a pair of 
objects be correlated by to the isometry $g_i$ at an angle 
$\gamma$, is 
\begin{eqnarray} 
\widehat{{\cal P}}^{g_i}(\gamma) 
= P_i \, \frac{\Theta(\gamma - \gamma^{min}_{g_i}) \,
\Theta(\gamma^{max}_{g_i} - \gamma)}{\Delta \gamma_{g_i}^{}}\, ,
\end{eqnarray}
where $\Delta \gamma_{g_i}^{} \equiv 
\gamma^{max}_{g_i} - \gamma^{min}_{g_i}$. 
Thus, the topological signatures coming from the isometries 
$g_i$ contribute to the EPASH (i.e. $\Phi_{expected}^{g_i}$, 
see Eq.~\ref{def_expected}) as 
\begin{eqnarray} \label{topo_sign}
\Phi_{expected}^{g_i} \equiv \, \frac{1}{\delta \gamma}
\int \widehat{{\cal P}}^{g_i}(\gamma)d\gamma 
= \frac{P_i}{\delta \gamma}\, .
\end{eqnarray}
Therefore, the angular correlations due to Clifford 
translational isometries appear in the PASHs as triangular 
peaks with basis $\Delta \gamma_{g_i}^{}$ and height 
$P_i / \delta \gamma$.

The effects of the non-translational isometries can also be
calculated using Eq.~(\ref{Isometry_def}). However, in 
such a case, the integrals are cumbersome to handle, so we 
prefer to show their effect in MPASHs with plentiful numerical 
simulations.
For this reason, we decided to work with the $G_6$ manifold,
which does not present Clifford translations in the histograms
whenever $R/L \leq \sqrt{2}/2$, $L$ being the arista of the
cube representing the FP.

\section{Numerical simulations}

We performed numerical simulations aiming to reveal topological 
signatures due to the isometries of the compact flat manifods 
$T^3, \,T_{\pi}$ and $G_6$ for full 
sky and for polar-cap simulated catalogs. 

It is dif\/ficult sometimes to recognize a topological signature 
in a single PASH due to the presence of statistical noise.
As a matter of fact, we found that topological signatures due 
to Clifford translations appear in PASHs as small triangular 
peaks whose heights are inversely proportional to the number 
of objects in a given catalog (see Eq.~\ref{TI}).
We saw that it is convenient to divide a catalog of $N$ objects 
in $K$ sub-catalogs of $\left< N \right> = N/K$ objects, 
where the objects in a sub-catalog share common physical properties,
and then perform the MPASH of the $K$ resulting PASHs. 
By doing this, we magnify in two ways the presence of the 
topological signatures: 
reducing the number of objects per catalog, 
$N \rightarrow \left< N \right>$, the height of the peak 
increases by a factor $N / \left< N \right> = K$;
and obtaining $K$ comparable catalogs we can perform the 
MPASH to reduce the noise by a factor $1/\sqrt{K}$. 
The net result is to increase the signal-to-noise ratio 
(SNR) of the topological signatures by a factor $\sqrt{K}$.
This explains why the strategy of dividing in sub-catalogs 
the original catalog with a large $N$, and performing 
the MPASH works better than considering just the PASH of 
the original catalog.
For 3D catalogs, that is, when the 3 spatial 
coordinates of the objects are provided, the 
Collecting Correlated Pairs method 
(Uzan et al. \cite{ULL}) is suitable to enhance the SNR 
for the signatures originated by the holonomies 
of the 3-space ${\cal M}_k$.

In the simulations presented in this section we assume 
that the FP is a cube of side $L=1$, and the origin of 
coordinates is in the center of this cube. 
Moreover, for $T^3$ and $T_{\pi}$, we consider 
$R \equiv R_{+} = 1, \, R_{-} = 0.997$, i.e. 
$\Delta R = 0.003$. 
The case of $G_6$ is doubly special: firstly, in order to 
avoid the presence of Clifford translations appearing at 
$\lambda = \sqrt{2}/2$, we assume $R = 0.7$; 
secondly, because the non-translational isometries have a 
very small signature in the MPASH, we assume $R_{-} = 0.69$, 
i.e. $\Delta R = 0.01$. 

In order to normalize and compare the results appearing in 
the dif\/ferent MPASHs shown below, we fix the values of 
$m$, and consequently that of $\delta \gamma$, in all of them.
Thus, we consider that the interval $[0,\pi]$ is divided in
$m = 400$ bins, each one of width 
$\delta \gamma = \pi/m = 0.00785$.

In general, to clearly reveal the signatures in the MPASH, 
a good SNR is of the order of 10; in particular, regarding 
Clifford translational isometries, this implies that the 
number of catalogs should be such that 
\begin{eqnarray}
\frac{P_i}{\delta \gamma} \, \stackrel{>}{_{\sim}} 
\, 10\, \sigma \, .
\end{eqnarray}
Thus, using the expression for $P_i$, Eq.~(\ref{P_i}), 
and the numerical values for the parameters involved, we 
obtain that, for $K \stackrel{>}{_{\sim}} 10^4$, the 
topological signatures due to Clifford translations are 
clearly revealed in the MPASHs.

\subsection{MPASHs for full-sky maps}

In this section, we show how to generate full-sky catalogs 
in order to construct the MPASHs to reveal topological 
signatures in the angular distribution of cosmic objects. 
We assume in all the simulations presented here that the FP 
is a cube of side $L = 1$, centered at the origin of 
coordinates. 
To construct a PASH through a numerical simulation, we first 
randomly generate a set of three Euclidean coordinates 
$x,y,z \in [-0.5,0.5]$ for $N_{seeds}$ objects located 
inside the FP.

In order to be sure that the distribution generated in this 
way is uniform, we test the randomness of the simulation 
using the Pair Separation Histograms for the simply-connected 
case ${\cal R}^3$, according to the theoretical relationship 
expected for such a case given by Bernui \& Teixeira (\cite{BT}).
After that, these $N_{seeds}$ objects are mapped to the whole 
ball ${\cal B}_R$ via the translational isometries 
$g_i \rightarrow \lambda_i = 1, \sqrt{2}, \sqrt{3}, \dots$
of \,$T^3$, giving rise to $N_{T}$ objects (including the 
$N_{seeds}$ objects) in ${\cal B}_R$.
Then, we select the $N$ objects located inside the thin 
shell, i.e. those objects with radial distance satisfying 
$R_{-} \leq r_i \leq R, \, i=1,...,N$. 
We project them on the involving sphere ${\cal S}_R$ and 
calculate all the angular distances between the $N\,(N-1)/2$ 
pairs of objects. 

\begin{figure} 
\includegraphics[width=9cm]{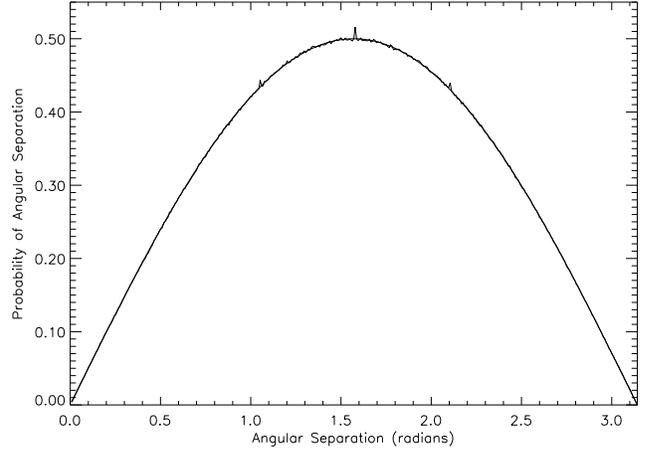}
\caption{MPASH together with EPASH for $K=10^4$ full-sky
simulated catalogs in $T^3$ (the FP is a cube of side 1); 
$\left< N \right>\, \simeq 100$ objects.}
\label{figure10}
\end{figure}

\begin{figure} 
\includegraphics[width=9cm]{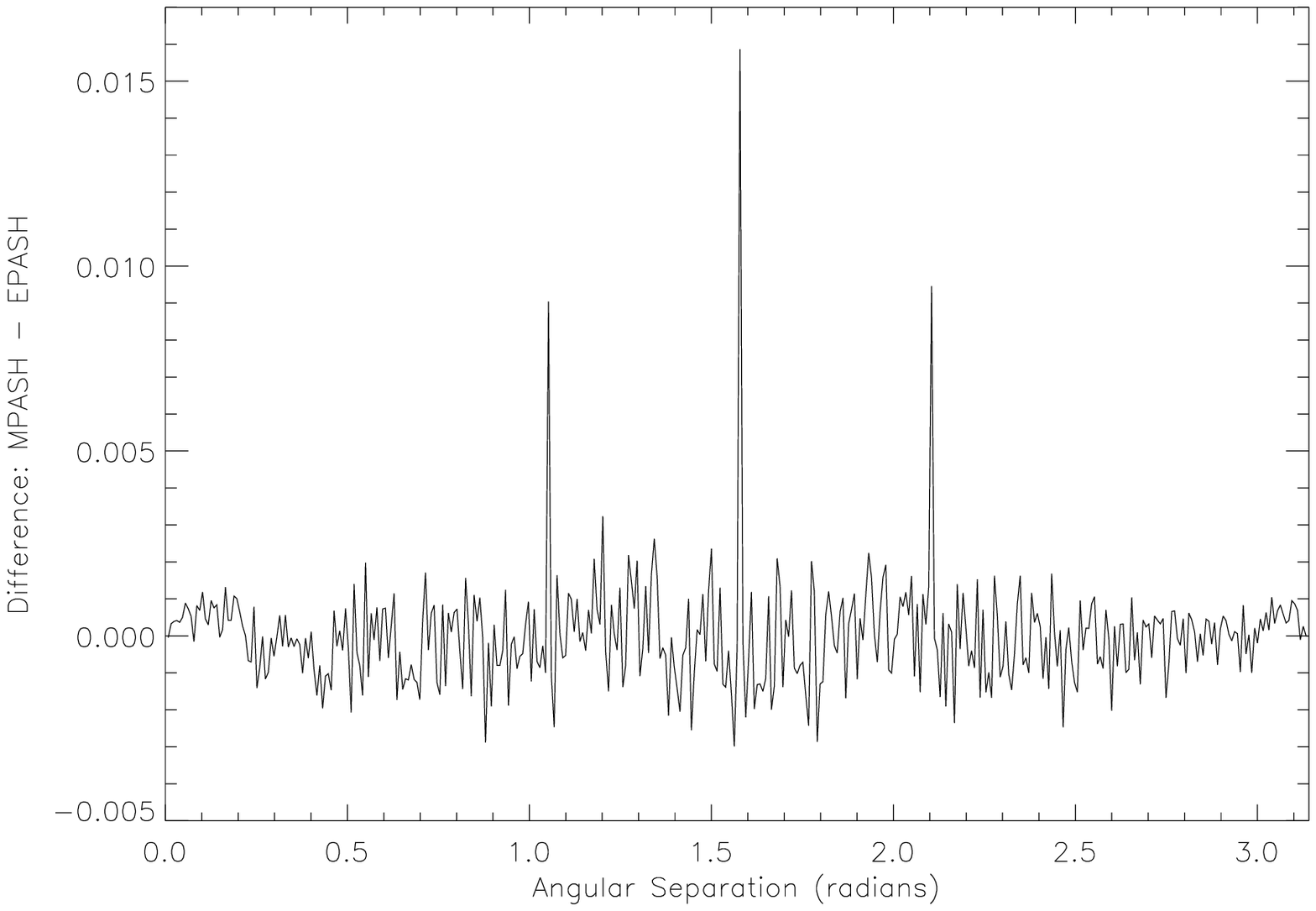}
\caption{Dif\/ference between MPASH and EPASH corresponding 
to Fig.~\ref{figure10}.}
\label{figure11}
\end{figure}

We then count how many of these angular distances are in 
the interval $J_i, \, i = 1,2, \dots , 400\,$. 
Finally, we normalize the PASH and plot the number of 
distances versus the angular distance $\gamma$. 
For the simulations showed in Figs.~\ref{figure10} 
and~\ref{figure11}, where we consider the case 
${\cal M}_0 = T^3$, we use $R = 1,\, R_{-} = 0.997$. 
In this case, due to the ratio between the volumes 
of ${\cal B}_R$ ($=4\pi/3$) and $T^3$ ($=1$) we should 
obtain images corresponding to the first three Clifford 
translations: $\lambda_1 = 1$, $\lambda_2=\sqrt{2}$, and
$\lambda_3=\sqrt{3}$.
As a consequence of this, the MPASH presents angular 
correlations that are approximately located at 
$\gamma^{min}_{g_i} \simeq 1.047,\, 1.571,\, 2.094$ radians, 
respectively, with the corresponding peak heights 
$P_i/\delta\gamma \sim 1.14 \times 10^{-2},\, 
1.6 \times10^{-2}$, and $0.9 \times 10^{-2}$.

\begin{figure} 
\includegraphics[width=9cm]{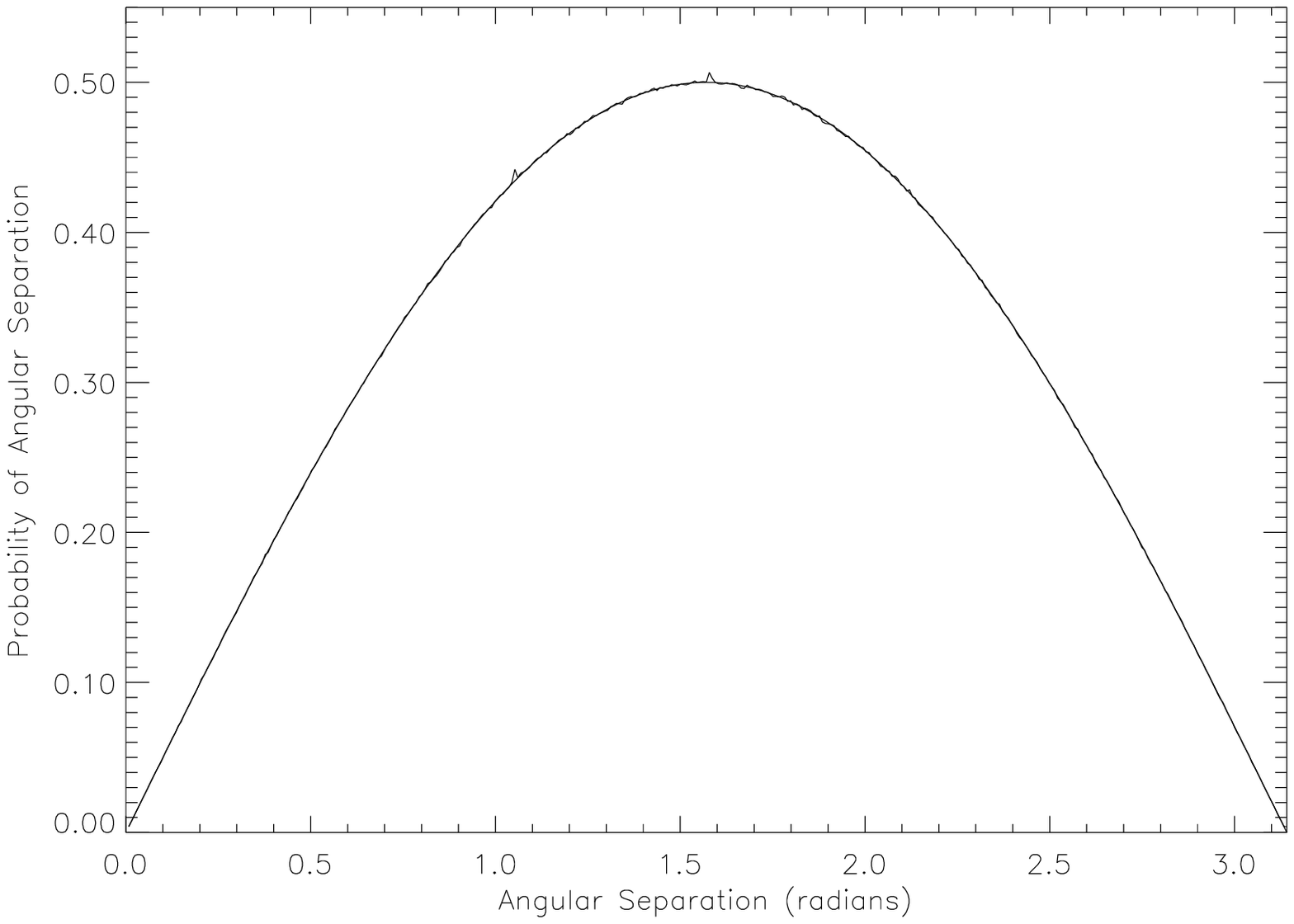}
\caption{MPASH together with EPASH for $K=2 \times 10^4$ 
full-sky simulated catalogs in $T_{\pi}$ (the FP is a cube 
of side 1), $\left< N \right>\, \simeq 100$ objects.}
\label{figure12}
\end{figure}

\begin{figure} 
\includegraphics[width=9cm]{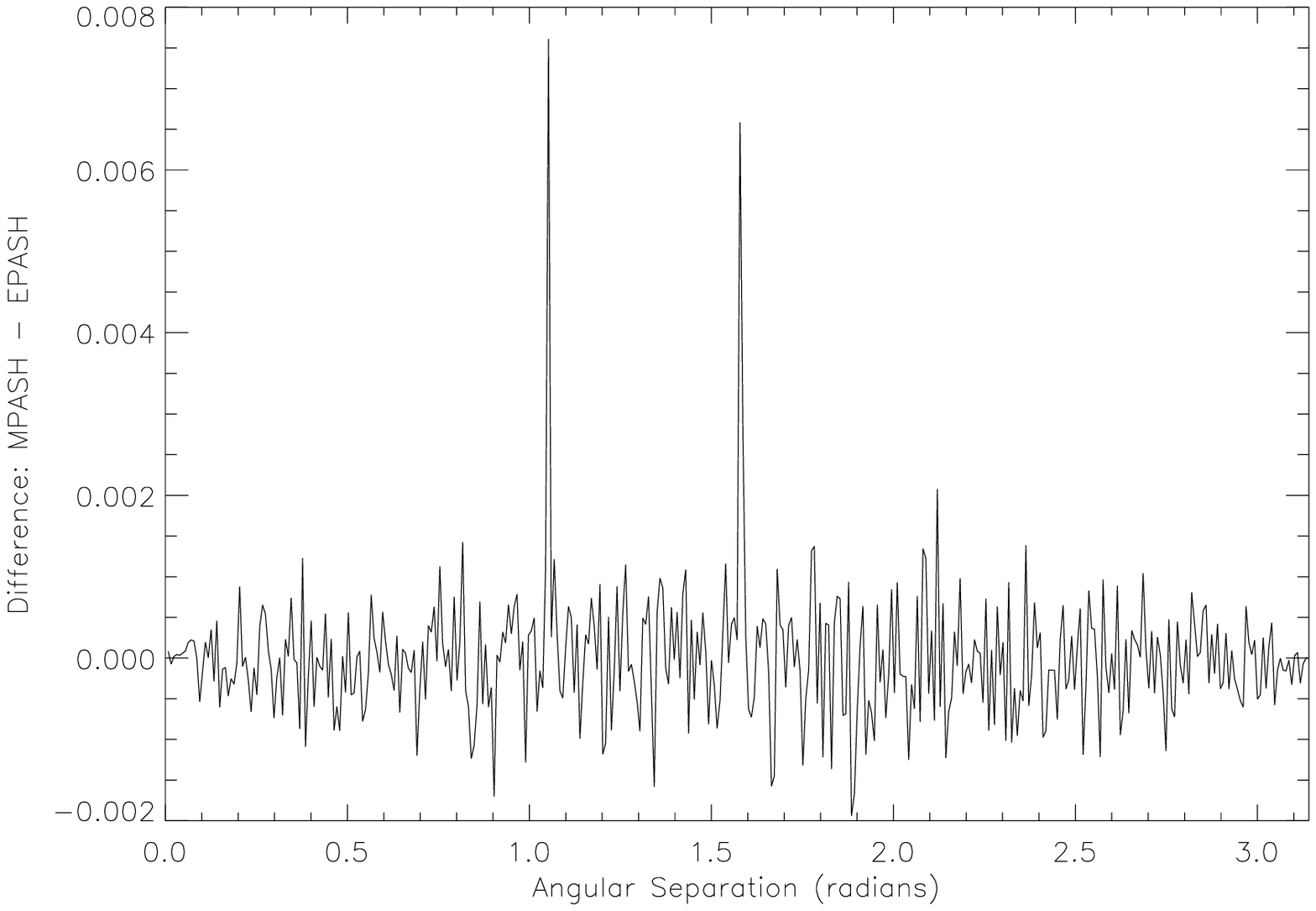}
\caption{Dif\/ference between MPASH and EPASH corresponding
to Fig.~\ref{figure12}.}
\label{figure13}
\end{figure}

For the simulation showed in Fig.~\ref{figure12}, where we 
study the case ${\cal M}_0 = T_{\pi}$, we also used 
$R = 1,\, R_{-} = 0.997$.
However, in this case, the Clifford translations contributing 
to the topological angular correlations studied in the 
previous section are just two: 
$\lambda_1=1$, $\lambda_2=\sqrt{2}$, 
which generate only the images $\nu_1 = 4,\, \nu_2 = 4$, 
respectively. 
According to this, the heights of the 2 peaks appearing in 
Fig.~\ref{figure12} are 
$P_i/\delta\gamma \sim 7.6 \times10^{-3},\, 5.4 \times10^{-3}$, 
and are approximately located at 
$\gamma^{min}_{g_i} \simeq 1.047,\, 1.571$ radians, 
for $i=1,2$ respectively. 
In fact, as we can observe in Fig.~\ref{figure13}, there 
is no peak at $\gamma^{min}_{g_i} \simeq 2.094$ because there 
is no translational isometry at $\lambda=\sqrt{3}$ in 
$T_{\pi}$.

\begin{figure} 
\includegraphics[width=9cm]{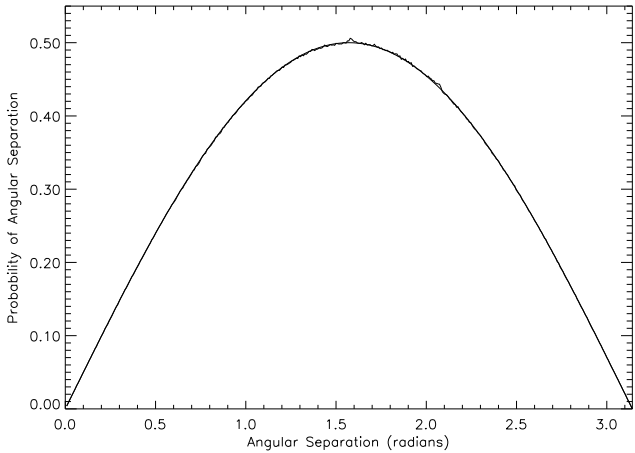}
\caption{MPASH together with EPASH for $K=4 \times 10^5$ 
full-sky simulated catalogs with objects in $G_6$ (the FP 
is a cube of side 1), $R = 0.7,\, R_{-}= 0.69$ (i.e. 
$\Delta R = 0.01$), $\left< N \right>\, \simeq\, 20$ objects.}
\label{figure14}
\end{figure}

\begin{figure} 
\includegraphics[width=9cm]{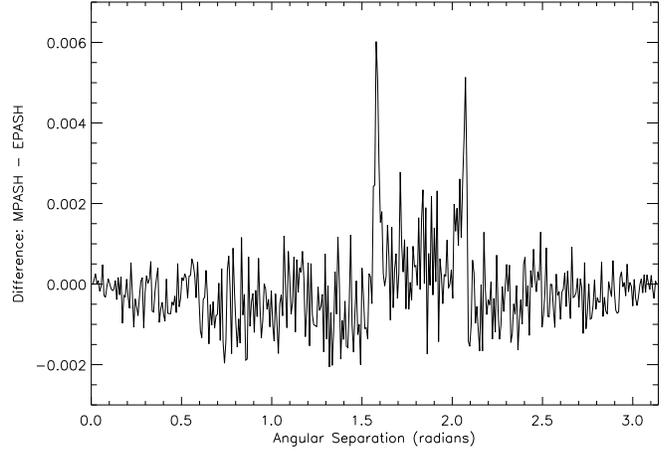}
\caption{Dif\/ference between MPASH and EPASH corresponding
to Fig.~\ref{figure14}.}
\label{figure15}
\end{figure}

In Fig.~\ref{figure14} we present the MPASH and the EPASH 
for the case of the 3-space ${\cal M}_0 = G_6$, and in 
Fig.~\ref{figure15} we show the corresponding dif\/ference 
between them. 
In this case, we assumed again that the FP is a cube of 
side 1, $R = 0.7,\, R_{-}= 0.69$, i.e. 
$\Delta R = 0.01$, $\left< N \right>\, \simeq\, 20$ objects, 
and $K = 4 \times 10^5$.

Comparing Figs.~\ref{figure11} and~\ref{figure13} with 
Fig.~\ref{figure15} we observe that the heights of the peaks 
are much smaller in the case of screw-motion isometries than 
in the case of Clifford translation isometries.
However, our simulations show that with a suitable number of 
catalogs, the screw-motions, as well as the pure translational 
isometries, are revealed in the MPASHs as small peaks.

\subsection{MPASHs for polar-cap maps}

We present numerical simulations in which the objects are 
located in a thin spherical shell of thickness $\Delta R$; 
the radial distance of the object to the origin of coordinates 
is $R \leq r \leq R_{-}$, $\theta_0 = 70^{\circ}$. 
We performed these calculations for both $T^3$ and $T_{\pi}$
3-spaces. 

A noticeable fact, shown in 
Figs.~\ref{figure11},~\ref{figure13}, and~\ref{figure15}  
corresponding to the numerical simulations done in the 
precedent sub-section, is that full-sky catalogs produce 
small distinguishable peaks in the MPASHs when the isometries 
are Clifford translations (which appears only in Euclidean 
and spherical geometries; for details see, e.g., Gomero et al. 
\cite{GRT1}, \cite{GRT2}, \cite{GRT3}, \cite{GTRB}), and even 
smaller signatures for the non-translational isometries. 

For completeness we perform now numerical simulations in order 
to test whether these results are valid for partial-sky catalogs, 
namely for polar-cap catalogs, and we do this for the cases when 
the 3-space is $T^3$ and $T_{\pi}$. 
Observing Figs.~\ref{figure17} and~\ref{figure18}, we notice 
that the topological information does not disappear when we 
consider a suitable polar-cap catalog instead of a full-sky catalog. 

Regarding our simulations, we remark that the values assumed 
for $R$ and $L$ means that the observable Universe includes 
a small number of multiple images, generated by the isometries 
of the manifold, for each seed-object present in the FP.
This assumption is within the limits imposed by COBE to the 
scale of the Small Universe considering a $T^3$ 
manifold (Stevens et al. \cite{SSS}; 
de Oliveira-Costa \& Smoot \cite{deOli-Smoot}).

\begin{figure} 
\includegraphics[width=9cm]{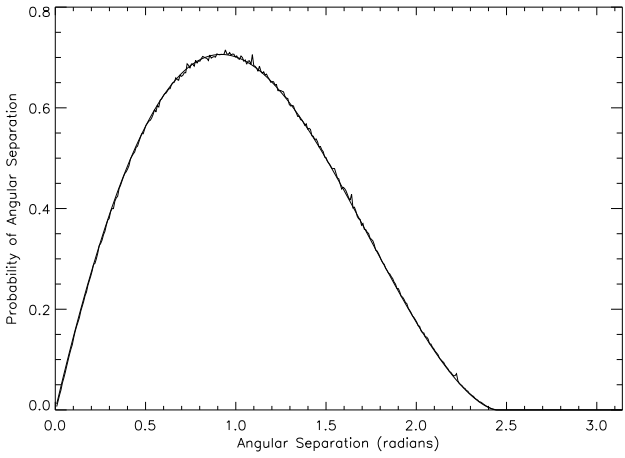}
\caption{MPASH together with EPASH for a polar-cap
catalog in $T^3$, with $\theta_0 = 70^{\circ}$, 
$\left< N \right> \,\simeq\, 40$ objects, $K = 4 \times 10^4$.}
\label{figure16}
\end{figure}

\begin{figure} 
\includegraphics[width=9cm]{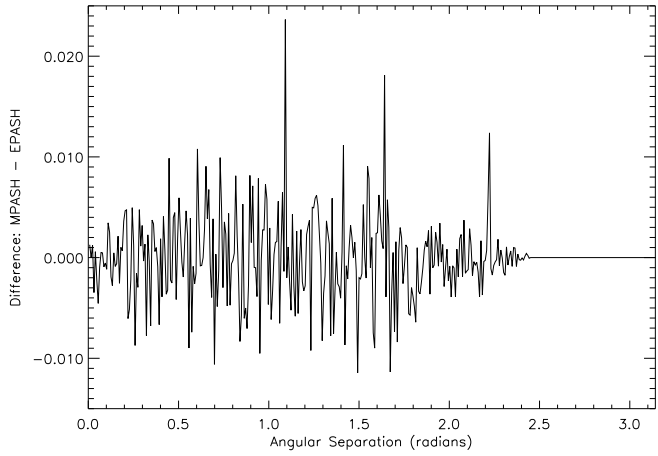}
\caption{Dif\/ference between MPASH and EPASH plotted in 
Fig.~\ref{figure16}.}
\label{figure17}
\end{figure}

\begin{figure} 
\includegraphics[width=9cm]{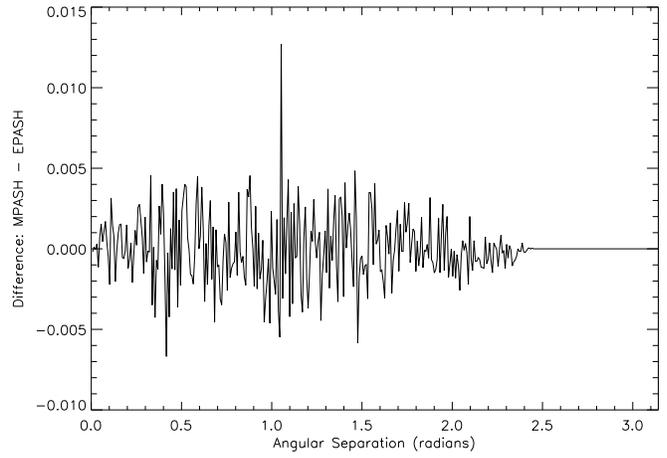}
\caption{Dif\/ference between MPASH and EPASH obtained for 
a polar-cap catalog (with $\theta_0 = 70^{\circ}$) for the 
case where the 3-space is $T_{\pi}$; 
$\left< N \right> \,\simeq\,40$ objects and $K = 4 \times 10^4$. 
The axis of symmetry of the polar-cap coincides with the axis 
of the screw-motion isometry.}
\label{figure18}
\end{figure}

\section{Discussion and conclusions}

We presented a method to study the angular distribution of 
cosmic objects. 
It has the great advantage of being independent of cosmological 
models or parameters. 
Through the analyses of the PASHs, we investigated the 
effect that dif\/ferent topological properties, termed 
Euclidean isometries, produce in the angular distributions 
of objects. We performed these analyses in full-sky as well 
as in polar-cap catalogs. 

We concentrate on the analysis of objects located in a thin 
spherical shell, considering only their angular positions in 
the celestial sphere in order to explore CMBR data. 
We have considered four dif\/ferent cases regarding the 3-space 
where the shell is embedded: the simply-connected ${\cal R}^3$, 
and the three topological\/ly dif\/ferent multiply-connected 
flat 3-spaces: $T^3$, $T_{\pi}$, and $G_6$. 

In the simply-connected case, topological angular correlations 
are clearly absent. This is evinced by plotting the 
dif\/ference between the MPASH and the EPASH, where only 
statistical fluctuations appear, as seen in Figs.~\ref{figure5} 
to~\ref{figure9}, independently of the number of simulated 
catalogs used to produce the MPASH. 
For the multiply-connected 3-spaces analyzed, angular 
correlations due to the isometries involved in each specific 
case indeed appear. 
Thus, the topological imprints corresponding to translational 
isometries manifest themselves as small and independent 
peaks, i.e., one peak for each isometry.
On the other hand, the signatures corresponding to the 
non-translational isometries (which is the case for the 
$G_6$ manifold) appear as a less intense but extended angular 
correlation: from $\gamma^{min}_{g_2} \simeq 1.571$ radians to 
$\gamma^{min}_{g_3} \simeq 2.094$, as can be seen in 
Fig.~\ref{figure15}.

It is also important to notice that the topological information 
of multiply connected 3-spaces does not disappear when we consider 
a suitable polar-cap catalog instead of a full-sky catalog, 
that is, provided that such polar-cap is suf\/ficiently large and 
conveniently oriented in the sky to map multiple images. 
This fact is clearly observed in Figs.~\ref{figure17} 
and~\ref{figure18}. 
This is a very useful result because in general astronomical 
data are presented in catalogs covering only partial regions 
of the sky.
In order to improve the SNR and reveal small topological 
imprints in the angular distribution of cosmic objects,
the strategy is to divide the original catalog in several 
sub-catalogs and then perform the MPASH. 
It could be also useful to divide the full-sky data in a set 
of antipodal polar-caps and compare them to similar sky patches 
in a statistically isotropic Universe (Bernui et al. \cite{BVF}).

We have shown that a topological signature, no 
matter how small it is, can always be revealed whenever 
one achieves a suitable SNR in the dif\/ference between the 
MPASH and the EPASH.
As shown in our simulations, the small topological signatures 
of $T^3$ and $T_{\pi}$ are clearly revealed when averaging 
$K=4 \times 10^{4}$ catalogs with $\langle N \rangle \simeq 40$ 
objects each, from an original catalog of 
$1.6 \times 10^{6}$ cosmic objects.
For $G_6$, the number of cosmic objects should be five times 
larger. 
For the case of CMBR maps which are always contaminated by
Galactic foregrounds, the polar-cap analysis described in 
this paper would be useful to reveal possible angular 
correlations in these data sets.

\begin{acknowledgements}

We thank A.F.F. Teixeira, G.I. Gomero, M.J. Rebou\c{c}as, 
and C.A. Wuensche for many fruitful discussions. 
A.B. thanks CNPq (PCI/DTI\/7B fellowship) and TWAS for the 
partial financial support, and the hospitality of the Centro 
Brasileiro de Pesquisas F\'{\i}sicas visited under the 
TWAS--Associateship Scheme at Centres of Excellence in the 
South. 
T.V. was partially supported by CNPq grant 302266/88-7-FA 
and FAPESP grant 00/06770-2.      
We also thank the anonymous referee for the
valuable comments on the manuscript.

\end{acknowledgements}

\end{document}